# Interaction and Student Dropout in Massive Open Online Courses

Markus Harju
Biomimetics and Intelligent Systems Group,
University of Oulu, Finland
email: markus.harju@oulu.fi

Teemu Leppänen
Center for Ubiquitous Computing,
University of Oulu, Finland
email: teemu.leppanen@oulu.fi

Ilkka Virtanen
Ionosphere Research Unit,
University of Oulu, Finland
email: ilkka.virtanen@oulu.fi

*Abstract*— Massive Open Online Courses (MOOC) are seen as a next step in distance online learning. In the MOOC vision, large numbers of students can access the course content over the Internet and complete courses at their own pace while interacting with their peers and instructors online. Despite the initial enthusiasm about MOOCs, large number of students were observed dropping out of the online courses. In this paper, we pinpoint the reasons behind the high student dropout rate and discuss how the interaction capabilities of MOOCs contributed towards the low completion rate.

## I. INTRODUCTION

Massive Open Online Courses (MOOCs) are rather new learning platforms in higher education. The idea of MOOCs is to make education freely available for anyone with an internet access in the form of free online courses. For example, a MOOC may consist of online video lectures, various types of assessments, and a discussion forum. Proponents of MOOCs claim that they will revolutionize the higher education by means of offering courses by the world leading experts for anyone willing to participate.

While top universities are behind several MOOCs, the courses may not belong to the curricula of the university students. The universities do not provide credits for all participants who complete the course, but the MOOC providers may grant their own certificates. However, the American Council on Education, an organization representing over 1800 colleges and universities, has recommended college credits to be granted for completing certain MOOCs (Ebben and Murphy, 2014). In Finland, the University of Helsinki offers a possibility to get university credits and to gain the right to study computer science in the university for students who successfully complete their "Introduction to Programming" MOOCs (Programming MOOC 2018).

The term "MOOC" was first used in connection with a course Connectivism and Connective Knowledge by George Siemens and Stephen Downes at the University of Manitoba in 2008, which had 2200 registered online students (Fini, 2009 ; Mackness et al., 2010). In 2011 professors Norvig and Thrun from the Stanford University attracted over 160 000 registered students from over 190 countries in their Introduction to AI course (Rodrigues, 2012). This truly massive course (in terms of number of students) led Thrun to inaugurate Udacity and its business model. A MOOC hype was quickly launched, and several commercial and non-profit organizations started to offer MOOCs on various topics. In fact, The New York Times even declared 2012 as "The Year Of the MOOC" (Pappano, 2012). As of today, the notable providers of MOOCs include Coursera, edX, Udacity and others.

The massive amount of students and openness to everyone leads to a huge diversity of the course participants, including the motivations behind the course registration, and creates challenges in pedagogical practices and interaction. A feature distinguishing MOOCs from traditional university courses is the large number of "dropout" students, who register for the course but do not complete it.

Because MOOCs are a relatively new form of education, also MOOC research is still emerging. For example, in 2014 Ebben and Murphy (2014) were able to identify only 25 peer-reviewed scientific articles about MOOCs, the first one published in 2008. The fast evolution of the field is reflected by the fact that the authors already divided the only six years and 25 papers long history of MOOC research into two distinct phases, the cMOOC and the xMOOC phase (see definitions below). Two years later, Veletsianos and Shepherdson (2016) were able to identify 183 papers published in 2013-2015.

The focus of this study is to find out the reasons for high dropout rate in MOOC courses and to see how the interaction capabilities of MOOC course execution had a role.

## II. MASSIVE OPEN ONLINE COURSES

Hoy (2014, pp. 85-86) defines MOOC as "an online course that anyone can participate, usually free of charge. MOOCs are made of short video lectures combined with automatically graded tests and online forums where participants can discuss the material or get help". While all MOOCs share the idea of free online access for anyone with a computer and a network connection available, there are some significant differences on how the courses are arranged. To this end, MOOCs are often categorized into cMOOCs and xMOOCs. Connectivist (or constructivist) MOOCs (cMOOCs) are built on human agency, user participation and creativity via connections provided by online technologies (Ebben and Murphy, 2014). Extended MOOCs (xMOOCs) have more traditional course structure as they are content-based and centralized (Margaryan et al., 2015).

The very first MOOC, Connectivism and Connective Knowledge 2008 (later "CCK08") course was a cMOOC, based on the theory of connectivism by the course organizers (Siemens, 2005 ; Downes, 2012). Their intention was to enable the participants to both engage with the theory of connectivism and to experience its principles in practice (Mackness, 2010). The theory of connectivism emphasizes that learning occurs in networks rather than by individuals. According to Downes (2012, p. 9), "Connectivism is the thesis that knowledge is distributed across a network of connections, and therefore that learning consists of the ability to construct and traverse those networks". A variety of technological tools for interacting with other students were used in CCK08. While participation required only blog and concept maps, more than 12 different tools (such as mailing list, Moodle, Wiki pages, web conferencing and social media platforms) were used during the course (Fini, 2009).

Today MOOCs are offered by several large providers, the leading ones being Coursera, Udacity and EdX. These courses are typically xMOOCs, in which the organizers have prepared the course material and one can complete the course with relatively little, or without any, interaction with the other students. Typical course material contains video lectures and computer assignments, which may also facilitate course participation in one's own pace. Interaction and collaborative learning are supported by means of discussion forums. As described by Rodriguez (2013, p. 71), the xMOOCs "rely primarily on information transmission, computer-marked assignments and peer assessment."

Studying in a MOOC has many similarities with the flipped classroom learning: the course material is available online and the students are expected to study independently. Brahimi and Sarirete (2015) even state that MOOCs follow the flipped classroom model. However, the complete lack of contact teaching makes MOOCs different from many university courses which follow the flipped classroom model.

The large MOOC providers can also monitor student behaviour in their courses. The course organizers record how users use the provided user interface, i.e. "clickstream data", which is a widely used method in MOOC research as it is easily available. Also network surveys can be easily arranged among the course participants. While these data are certainly valuable, following student behaviour with such accuracy is impossible in traditional courses, e.g. the results by Veletsianos and Shepherdson (2016) suggest that the automatically collected and survey data dominate the MOOC research, while "very few studies were informed by methods traditionally associated with qualitative research approaches". The authors also claim that "the researchers have favored a quantitative, if not positivist approach to the conduct of MOOC research".

*A. Pedagogical practices for MOOCs*

The idea of a practically unlimited number of participants in a MOOC creates an obvious challenge for pedagogical practices in MOOCs. The teachers cannot have direct interaction with each individual student and they cannot e.g. grade exams from all participants. The latter imposes restrictions on evaluation methods and even course topics. Moreover, the students cannot interact with the majority of their peers, either.

Abeer (2014) notes that the role of the lecturer in a MOOC differs from his role in traditional open courses in the need to cope with a massive number of students from different cultures, in the need to address technical problems related to inadequate technologies in different countries, and using a variety of teaching methods in order to make online learning more interactive and interesting.

Margaryan et al. (2015) studied the pedagogical quality of 76 randomly selected MOOCs, including both cMOOCs and xMOOCs. They studied if the courses followed the First Principles of Instruction by Merril (2002). According to the principles, learning activities should be problem centred and they should contain activation, demonstration, application and integration. In addition, the learning should support collective knowledge, collaboration, differentiation (according to students' needs), authentic resources and feedback. The authors found that, in light of these principles, MOOC can be considered generally low quality. When each course was scored and the highest possible score was 72, the best score was 28 and median only 9. Interestingly, even the selected cMOOCs contained only very limited amount of collaborative learning.

In addition to the instructional principles, Margaryan et al. (2015) studied the quality of organization and presentation of the selected MOOCs. They examined whether or not the courses had measurable learning objectives, learning outcomes, well-organized material, and clear course requirements and descriptions. The majority of the courses did not specify learning objectives and expected outcomes nor they were measurable. However, the majority of the courses had good quality in the organization of the course material and provided clear course descriptions and requirements.

Toven-Lindsay et al. (2015) studied pedagogical practices in MOOCs. They explored "the range of pedagogical tools used in 24 MOOCs, including the epistemological and social dimensions of instruction, to consider the extent to which these courses provide students with high-quality, collaborative learning experiences." The course topics were diverse, including e.g. biology, computer graphics, nursing, poetry and law. The authors found that the pedagogical practices followed mainly the objectivist-individual approach, where knowledge is transferred one-directionally from the teacher to the students. The authors point out that this approach hardly supports engaged learning, which constructivism and critical pedagogy consider "as the most valuable to encouraging both active learning and active democratic citizenship" (Toven-Lindsay et al. 2015, p. 11). The latter refers to the MOOC proponents' view of MOOCs as democratizing power. The authors conclude that the study raises some important concerns about a potential "MOOC revolution" in higher education.

III. REASONS FOR STUDENT DROPOUT

A striking feature of MOOCs is that a vast majority of students registered for a course never complete it. This may cause problems to a MOOC provider, often a business venture of some kind, since the MOOC platform must be financed in order to secure its long-term growth (Nawrot and Doucet, 2014). The high number of dropouts may also raise questions about the quality and usefulness of the MOOC education in general.

The central question in this regard is how dropping out is defined. If one uses the traditional definition, where all registered users are counted as participants and anyone who has not successfully completed the whole course is a dropout, many MOOCs will see dramatic dropout rates. A dropout rate of 90 percent or more are often mentioned in literature (Rieber, 2017 ; Ebben and Murphy, 2014 ; Brahimi and Sarirete, 2015, and references therein). A general completion rate of 15 percent, based on a collection of data from various sources, is reported by Jordan (2015). This is an alarming level compared with traditional university courses and makes MOOCs stand out as misfits in the eyes of university deans and rectors.

Because MOOCs are very different from traditional university courses in the sense that the students are not taking them as a part of a degree program, also their reasons for dropping out, as well as their initial motivation to register, may be different. A student who drops out in the middle of the course may never have intended to complete the course, in the first place. The hasty conclusion that the MOOC has failed because students are dropping out may thus not be justified, but one needs to study the reasons behind dropping out more carefully. According to Fini (2009), participants who do not complete a course cannot be considered as dropouts at all, because they may have reached their personal learning goals.

The high dropout rates are well-known phenomena in distance education, already before the MOOCs. Lack of time being the most common reason for dropping out both in some earlier distance education courses and in the CCK08 MOOC (Fini, 2008). In the survey, with 83 responses (Fini, 2008), "Lack of time" was selected 39 times, while six other reasons for not completing the CCK08 MOOC were selected by only 1-3 students each and "other reasons" was selected by 16 students. 17 students did not answer this question.

The students competence and the quality of the MOOC also play a role. Students, who lack the basic competencies, even if they study a well-designed MOOC, will probably dropout throughout the course. Similarly, students with high competencies, learning in an ill-structured MOOC, will probably fail to finish the course (Abeer, 2014).

deBoer et al. (2014) studied student behaviour in MOOCs using data from the first MOOC offered by MIT, i.e. Circuits and Electronics. The course attracted 154 763 registered students, the majority of whom registered before or at the course start date. However, the registration was open until the end of the course and new registrations were recorded until the very end of the course. Only 70 % of the registered students clicked at least once in the actual course, 50 % clicked on a lecture video, 20 % attempted a homework problem and 8 % posted on a discussion forum. Based on these numbers, the authors conclude that (p. 77) "if enrollment is interpreted as the number of people who make an informed commitment to complete a course, the number of total course registrations in a MOOC is a naive operationalization at best". The authors suggest that the variables enrollment, participation, curriculum and achievement should be reconceptualized to make them a better match with the reality of MOOCs. They do not provide clear alternative definitions, but outline that the concepts should reflect the different tracks and goals of individual students.

Similar opinions are presented by Rieber (2016), who notes that about 78% of registered participants do not subsequently participate in any way. The author made a survey in a statistics course with 5079 students, out of whom 1935 responded to the survey. About half of the respondents reported that they intended to complete the course while a quarter of the students intended to "fully participate in most of the course". Eventually, only 35% of the students who intended to complete it did so, while 19% of the students who did not intend to complete actually did. The main reasons for not completing the course despite the initial intention to do so were "I did not have enough time to complete the course" and "The time needed to complete the course was greater than I expected". Although Rieber recognizes the risk of offering an explanation for behaviour of participants who did not answer the survey, the "shopping theory" of deBoer (2014) is suggested. As the registration for a MOOC does not cost anything, the students may be registering on one MOOC after other when looking for a sufficient one, just like people visit several stores when trying to find a gift to purchase.

Both deBoer et al. (2014) and Rieber (2016) suggest that the completion rates should not be calculated based on the number of registered students, but only students who show at least some level of commitment to the course should be counted. deBoer et al. (2014) demonstrate that even a rather modest definition of commitment could lift the course completion from 5% close to 50%.

The majority of the dropouts in MOOC may be "harmless shoppers", in the sense that there was not necessarily any particular fault in the course leading to the dropout. However, the significant fraction of students who showed commitment and intended to complete the course but failed deserves better attention. Rieber (2017) found that (p. 1302) "(1) People who enroll in a MOOC with the declared intent to complete the MOOC are more likely to do so than those who do not declare this intent; and (2) Successfully completing initial course milestones contributes to the probability that people will complete the course regardless of their originally stated goals." The author states that student participation could be retained by means of including "low stakes" activities, in which the students will succeed with a high probability, throughout the course.

Kizilcec and Schneider (2015) conducted an online survey on 14 MOOCs offered by Stanford University. Topics of the courses included biology, computer science, material science, mathematics, medicine, political science, quantum physics, sociology, statistics and writing. The number of enrolled students varied between 3000 to 60000. The authors note that (p. 2) "Many learners interact with these courses in ways that would not be considered "successful" with respect to instructor-defined criteria of success. In contrast, these learners' behaviors would be considered normal, or even successful, in the context of engagement in online media, where user-driven behaviors are welcome and encouraged". A student may, for example, follow all lectures of a course, but skips assignments to save time. The student may have got the information he/she was originally looking for, but will be counted as a dropout in the statistics. The authors found that only 45% percent of course participants intended to gain a course certificate, which is close the 50% reported by Rieber (2016). On the other hand, student motivations included meeting new people in the course (25%) and taking a course with friends or colleagues (20%). Furthermore, 28% were motivated to improve their English. Simple curiosity to experience an online course was reported by 44% of the respondents.

Regarding course participation and success, Kizilcec and Schneider (2015) did not find significant correlation between the intention to earn a course certificate and actually gaining it. The only factor that was found to improve the probability of gaining a certificate was taking the course with a friend or colleague. The authors found that learners who considered the course content as relevant to their study or research were watching less video lectures and doing less assignments. The authors suggest that these students may be using the MOOC as a reference source, and have actually found the information they were looking for. Furthermore, these users would benefit if MOOC contents were divided into tagged modules to help in their reference-style use.

Nawrot and Doucet (2014) conducted an online survey (n=508). The respondents in this survey were recruited using an online crowdsourcing platform and they did not belong to any particular MOOC nor were their actual participation in a MOOC investigated. They found out that 68.9% of the respondents attributed "bad time management" to their withdrawal from MOOC learning. Other reasons for dropping out were related to the attractiveness and suitability factors of the MOOC.

## IV. Interaction in MOOC

The primary interaction tool provided by the MOOC platforms is often a discussion forum, where the students and instructors can read and post messages. The students may also write their personal blogs about course-related topics, etc. In addition, the students may be using many other tools outside the MOOC platform, such as Facebook, Twitter and Skype. While student interaction in the discussion forum can be easily studied by means of counting forum posts and clickstreams, etc, the type and amount of interaction with other tools may be poorly known.

Especially the teacher-student interaction in MOOCs may be very limited, because the huge number of students in a MOOC makes direct interaction with each student impossible. As reported by Toven-Lindsay et al. (2015), the connection between teachers and students tends to be one-directional transfer of information in video lectures etc., without much discussion between the teacher and the students.

According to Mackness et al., (2010, p. 268), "Ultimately the majority of asynchronous interaction took place in the Moodle forums and blogs, with participants interacting in either blogs or forums, or both" in the original CCK08 MOOC. However, a large number of blog posts were generated in a short time, and participants were consequently encouraged to interact from their blogs. Following all the discussions in a large course is impossible and (p. 272) "it seems that the larger the course, the more potential for interaction to degenerate into interference and noise." Students complained also about "tittle tattle", bad behavior, trolling, inability to follow the discussions, etc. Also the idea of openness and sharing caused challenges, because the necessary trust in between the numerous participants could not be easily created. The authors concluded that moderation, intervention to prevent negative behaviour, and communication about what is acceptable may be needed in MOOCs.

The effect of active interaction in a discussion forum has on the actual learning outcomes is not plain, either. Kizilcec and Schneider (2015) found that while the reported intention to earn a course certificate predicted more active course participation, including more active interaction in the discussion forum, these students were actually not more probable to actually gain the certificate than those who did not intent to earn one. However, a positive correlation between taking the course with friends and colleagues and earning a certificate was found. The students taking the course together with someone else were more engaged with the course material but less active in the discussion forum, possibly because they had the social interaction with the people they were taking the course with.

The benefits of taking the course with a friend hint about a way around the interaction issues, namely combining MOOCs with classroom teaching. Brahimi and Sarirete (2015, p.604) refer to what Bill Gates said in his speech in 2013: "The value of MOOCs comes when you use them to create hybrids that are the best of both worlds. Rather than having the instructor lectures during class and then send the students home with assignments, many instructors are now using MOOCs to flip the classroom''. However, this approach removes the scalability to an unlimited number of students and independence on time and place.

Fidalgo-Blanco et al. (2014) studied the "Applied Educational Innovation" MOOC with 6149 students. Their goal was to investigate how informal learning activities (e.g. interaction with classmates and social media tools) affected the learning results. They found out that there is a correlation, albeit weak, between informal learning and students' perceived results of learning. The researchers explain this phenomenon by the fact that the MOOC consisted of both formal and informal learning activities. However, the study revealed that the positive opinion of students' own learning is directly proportional to their participation in social interactions. We note that the research did not investigate the effect of informal learning activities and actual learning outcomes.

In a study about MOOC retention, Hone and el Said (2016) made a survey among 379 Egyptian students, who participated in a MOOC as an optional self-learning component in a university course. They were unexpectedly unable to observe an effect of student-student interaction, but the reported teacher-student interaction was found to have a positive effect on student retention. The authors speculate that good instructor-student interaction might be necessary for a good student-student interaction. The student-student interaction might have been insufficient because the instructors could not support and direct discussions. From free text answers, the authors make the interesting finding that those who completed the course did not mention interaction, but those who did not complete the course reported poor communication and the feeling of isolation. The authors conclude that the instructor-student interaction may become a limiting factor for very large MOOCs and that MOOC providers should look for better ways to provide appropriate human interaction in their courses.

## V. Discussion

MOOCs are a relatively new form of education, with great promise to bring free, high-quality education available to anyone, practically anywhere in the world. However, the experiences of using MOOCs are largely critical towards the expected benefits. The massive scale and lack of contact teaching introduce problems, which are not present in traditional university courses. Even recognizing the actual issues may by challenging, because student behavior traditionally considered as a failure may actually be exactly what the person was initially looking for. In particular, not completing a course may not be a failure in the MOOC world, where the students may register for courses only to see how they look like or only to pick some or one particular piece of information.

We have seen that despite the initial attractiveness of a MOOC, in many cases a high number of participants do not complete the course. While the student may have left the course because he/she had already found the necessary information, this may create a "fast food" type behavior in learning, in which goals or principles of deep learning are not reached.

As the number one reported factor leading to dropout is lack of time, student's time management skills come to play. MOOCs should be built in a way that supports students' good time management. Several self-regulation methods to improve the MOOC course achievements (N=331) were studied in (Kizilcec et. al. 2016). The methods suggested by successful learners were recommended. It was found out that just introducing these methods does not improve success, but it was suggested that the self-regulation methods should be deeply integrated into the course execution.

In an online setting, it is obvious that the teaching staff cannot create conditions similar to traditional classroom studying nor can the students get peer-to-peer support equally well. The lack of interaction and support have been recognized as negative sides of MOOCs by students who dropped out (Hone and el Said, 2016), giving a reason to believe that better interaction tools could have assisted in continuing in the course. Also, the finding that students studying together with friends and colleagues are more likely to complete the course (Kizilcec and Schneider, 2015), hints that interaction with peers may have helped these students.

However, good student-student interaction may be difficult to create in between students previously unknown to each other. The online discussions were partially problematic in the CCK08 MOOC (Mackness et al., 2010) and the need for instructors to direct the online discussions was speculated by Hone and el Said (2016). If the student-student interactions cannot be improved without direct help from the instructors, organizing very large MOOCs may become impractical.

In light of the existing research results, insufficient interaction with both the instructors and the other students may have an effect on the student dropout level in MOOCs. However, even the definition of a dropout is not at all clear in connection to MOOCs, and the available research is still limited. Even the limited available literature has been accused of being biased towards only a few research methods and towards a positivist attitude on MOOCs (Veletsianos and Shepherdson, 2016). Further research on interactive MOOCs is thus obviously needed.

ACKNOWLEDGMENT

This study was carried out during the "University pedagogy for teachers"-training program, organized by the Faculty of Education, University of Oulu, Finland. The authors would like to thank all their colleagues and instructors for the discussion regarding this study.

REFERENCES

Abeer, W., and Miri, B., Students' preferences and views about learning in a MOOC. *Procedia-Social and Behavioral Sciences*, 152, 318-323, 2014.

deBoer, J., Ho, A. D., Stump, G. S. and Breslow, L., Changing Course: Reconceptualizing Educational Variables for Massive Open Online Courses, *Educational Researcher*, 43 (2), 74-84, 2014.

Brahimi, T. and Sarirete, A., Learning outside the classroom through MOOCs, *Computers in Human Behavior*, 51, 604-609, 2015.

Downes, S., Connectivism and Connective Knowledge, 2012.

Ebben, M. and Murphy, J. S., Unpacking MOOC scholarly discourse: a review of nascent MOOC scholarship, *Learning, Media and Technology*, 39 (3), 328-345, 2014.

Fidalgo-Blanco, A., Sein-Echaluce, M. L., Garcia-Penalvo, F. J. and Escano, J. E., Improving the MOOC learning outcomes throughout informal learning activities, In: TEEM '14 Proceedings of the Second International Conference on Technological Ecosystems for Enhancing Multiculturality, 611-617, 2014.

Fini, A., The Technological Dimension of a Massive Open Online Course: The Case of the CCK08 Course Tools, *International Review of Research in Open and Distance Learning*, 10 (5), 2009.

Hone, K. S. and El Said, G. R., Exploring the factors affecting MOOC retention: A survey study, *Computers & Education*, 98, 157-168, 2016.

Hoy, M. B., MOOCs 101: An Introduction to Massive Open Online Courses. *Medical Reference Services Quarterly,* 33, 85-91, 2014.

Jordan, K., MOOC completion rates. The data. Retrieved from http://www.katyjordan.com/MOOCproject.html (accessed 16 March 2018).

Kizilcec, R. F., and Schneider, E., Motivation as a lens to understand online learners: Toward data-driven design with the OLEI scale. *ACM Trans. Comput.-Hum. Interact. 22* (2), article 6, 2015.

Kizilcec, R. F., Pérez-Sanagustín, M., and Maldonado, J. J., Recommending self-regulated learning strategies does not improve performance in a MOOC. In: Proceedings of the Third ACM Conference on Learning@ Scale, 101-104, 2016.

Mackness, J., Mak, S. and Williams, R., The ideals and reality of participating in a MOOC, In: Proceedings of the 7th International Conference on Networked Learning, 2010.

Margaryan, A., Bianco, M. and Littlejohn, A., Instructional quality of Massive Open Online Courses (MOOCs), *Computers & Education, 80*, 77-83, 2015.

Merrill, M. D., First principles of instruction, *Educational Technology Research and Development*, 50 (3), 43-59, 2002.

Nawrot, I. and Doucet, A., Building Engagement for MOOC Students: Introducing Support for Time Management on Online Learning Platforms, In: Companion Proceedings of the 23rd International Conference on World Wide Web Pages, 1077-1082, 2014.

Pappano, L. (2012). The Year of the MOOC, The New York Times, http://www.nytimes.com/2012/11/04/education/edlife/massive-open-online-courses-are-multiplying-at-a-rapid-pace.html (accessed 22 March 2018).

Programming MOOC 2018, http://mooc.fi/courses/2018/ohjelmoinnin-mooc/ (accessed 23 March 2018).

Rieber, L. P., Participation patterns in a massive open online course (MOOC) about statistics, *British Journal of Educational Technology*, 48 (6), 2017.

Rodriguez, C., MOOCs and the AI-Stanford like Courses: Two Successful and Distinct Course Formats for Massive Open Online Courses, http://www.eurodl.org/materials/contrib/2012/Rodriguez.htm (accessed 14 March 2018).

Rodriguez, O. (2013), The Concept of Openness Behind c and x-MOOCs, *OpenPraxis* 5 (1), 67–73, 2013.

Siemens, G., Connectivism: A Learning Theory for the Digital Age, *International Journal of Instructional Technology & Distance Learning*, 2 (1), 2005.

Toven-Lindsay, B., Rhoads, R. A. and Lozano, J. B., Virtually unlimited classrooms: Pedagogical practices in massive open online courses, *Internet and Higher Education,* 24, 1-12, 2015.

Veletsianos, G. and Shepherdson, P., A Systematic Analysis and Synthesis of the Empirical MOOC Literature Published in 2013–2015, *The International Review of Research in Open and Distributed Learning*, 17 (2), 2016.